\documentclass[12pt]{article}

\usepackage{amsfonts,amssymb,amsmath}

\newtheorem{theorem}{Theorem}[section]
\newtheorem{pro}[theorem]{Proposition}
\newtheorem{defin}[theorem]{Definition}

\newcommand{\bse}{\boldsymbol{e}}

\newcommand{\bsp}{\boldsymbol{p}}

\newcommand{\bst}{\boldsymbol{t}}

\newcommand{\bsR}{\boldsymbol{R}}

\newcommand{\rr}{{\mathbb R}}
\newcommand{\zz}{{\mathbb Z}}
\newcommand{\GG}{\mathcal{G}}
\newcommand{\LL}{\mathcal{L}}
\newcommand{\PP}{\mathcal{P}}
\newcommand{\HH}{\mathcal{H}}
\newcommand{\KK}{\mathcal{K}}
\newcommand{\MM}{\mathcal{M}}
\renewcommand{\SS}{\mathcal{S}}

\begin{document}

\title{An algorithm for the arithmetic classification of multilattices}
\author{Giuliana Indelicato \\ 
\small Dipartimento di Matematica
\\ \small Universit\`a di Torino
\\ \small  Via Carlo Alberto 10, I--10123 Torino, Italy}
\date{}
\maketitle

%\begin{abstract}

%A procedure for the construction and the classification of multilattices in arbitrary dimension is proposed. The algorithm allows to determine explicitly the location of the points of a multilattice given its space group, and to determine whether two multilattices are arithmetically equivalent. The algorithm is based on ideas from integer matrix theory, in particular  the reduction to the Smith normal form.
%Among the potential applications of this approach is  a software package that generalizes the International Tables of Crystallography, allowing the classification of complex crystalline structures and the determination of their space groups. Not last, it can be used to determine the symmetry of regular systems of points in 6D, with applications to the study of viral capsid structures.
%\end{abstract}

%
\begin{abstract}

A procedure for the construction and the classification of multilattices in arbitrary dimension is proposed. The algorithm allows to determine explicitly the location of the points of a multilattice given its space group, and to determine whether two multilattices are arithmetically equivalent. The algorithm is based on ideas from integer matrix theory, in particular  the reduction to the Smith normal form.
Among the applications of this procedure is  a software package that allows the classification of complex crystalline structures and the determination of their space groups. Also, it can be used to determine the symmetry of regular systems of points in high dimension, with applications to the study of quasicrystals and sets of points with noncrystallographic symmetry in low dimension, such as viral capsid structures.

\end{abstract}

\section{Introduction}

 A monoatomic $(N+1)$-lattice is a regular set of points in $\rr^n$ that is the superposition of $N+1$ identical Bravais lattices, and can be described by a reference (or skeletal) Bravais lattice and $N$ shift vectors that measure the location of $N$ points relative to a fixed point of the reference lattice (cf. e.g., \cite{pitteri-zanzotto}). 
 
 The symmetry of a multilattice is  determined by those symmetry operations of the point group of the skeletal lattice that act as  permutations (modulo the lattice) on the additional points (\cite{parry_04}, \cite{fadda_zanzotto_99}, \cite{fadda_zanzotto_01}, \cite{fadda_zanzotto_01_acta}, \cite{pitteri_zanzotto_2000}, \cite{pitteri_zanzotto_1998}). More precisely, a symmetry operation of a $(N+1)$-lattice is a triple $M=(M_j^i)$, $A=(A_\alpha^\beta)$, $T=(T_\alpha^i)$ such that, for every $\alpha=1,\dots,N$,
 \begin{equation}\label{master0}
\sum_{j=1}^n M^i_jP^j_\alpha=\sum_{\beta=1}^NP_\beta^iA^\beta_\alpha+T^i_\alpha,
 \end{equation}
 with $M$ a unimodular integer matrix which is an element of the lattice group of the skeletal lattice, $A$ an integer matrix that describes the permutation action of $(M_j^i)$ on the points of the multilattice, $T$  a matrix of integers representing a set of lattice translations, and $P^j_\alpha$ the components of the $\alpha$-th shift vector. To each triple $(M,A,T)$ 
 a matrix of the form  
 \begin{equation}\label{matricione0}
 \left(\begin{array}{ll}M&T\\0&A\end{array}\right),
 \end{equation}
 can be associated, and it turns out that the set of all triples that satisfy  (\ref{master0}) for a given set of shift vectors is a group under matrix multiplication, the symmetry group of the multilattice.
 
Denote by $\Gamma_{n,N}$ the group of all matrices of the form (\ref{matricione0}) for arbitrary $M$ unimodular integer, $A$ a linear representation of a permutation on the shifts, and $T$ an integer matrix: two $(N+1)$-lattices are arithmetically equivalent 
if their symmetry groups are conjugated in $\Gamma_{n,N}$. This notion of equivalence generalizes to multilattices the usual arithmetic classification of simple lattices in Bravais types. 

We refer to (\ref{master0}) as the master equation of the multilattice.  It can be used either to compute the shift vectors $P$, given the symmetry group, or to compute the symmetry group given the skeletal lattice and the shift vectors. 
 
In this work we describe an algorithm to solve  the master equation for any given symmetry of the skeletal lattice.  Precisely, we  assign a subgroup of the lattice group of the skeletal lattice, construct a permutation representation of this group, and solve the corresponding master equation, letting the integer translations be taken care of by simple bookkeeping. The procedure is based on ideas from integer matrix diagonalization (\cite{smith_1861}, \cite{gohberg}, \cite{havas_bohdan_majewski} , \cite{jaeger}, \cite {newman}, \cite {dumas}) and  yields automatically a single representative for each arithmetical equivalence class of multilattices.

Hence, this approach provides a basis for an automatic procedure for the classification of multilattices with arbitrary number of points, but is also yields a simple method to determine  regular sets of points in arbitrary dimensions. This sort of calculations is useful for instance when high-dimensional crystallography is used, via a projection approach, to study quasicrystals or sets of points with noncrystallographic symmetry, such as viral capsids that display icosahedral symmetry.

\section{Multilattices and the master equation}

\subsection{Multilattices}
In this section we introduce the notion of multilattice.  We first recall some basic definitions about simple lattices, also called Bravais lattices. Unless otherwise stated, we use the convention that the sum over repeated index is implied.

\begin{defin} A Bravais lattice, with basis $\{\bse_i\}_{i=1\dots n}$ (a basis of $\rr^n$) and origin  $Q_0\in\rr^n$, is the set of points  in $\rr^n$ defined  by
\begin{equation}
\LL=\LL(Q_0,\bse_i):=\{Q_0+m^i\bse_i \in \rr^n\; / \; m^i\in\zz \;\mbox{for}\; i=1\dots n    \}.
\end{equation}
\end{defin} 
Any other lattice basis $\{\bse_i'\}_{i=1\dots n}$ of $\LL$ is an integer linear combination of the vectors  $\bse_i$:
\begin{equation}
\bse_i':=M^j_i\bse_j\qquad  M=(M^j_i)\in GL(n,\zz)   
\end{equation}
with $GL(n,\zz)$ the group of integer unimodular matrices.
{\defin\label{point_group} The point group $\PP$ of a lattice $\LL$ is the group of orthogonal transformations whose action on a lattice basis corresponds to a change of basis of the lattice:
\begin{equation}
\label{point_group_eq}
\PP=\left\{
\bsR \in O(n)\; / \; 
\exists M=(M_i^j)\in GL(n,\zz) \; / \; \bsR \bse_i = M_i^j \bse_j\right\}
.
\end{equation}
}
\begin{defin}
The lattice group $\GG$ of $\LL$ is the subgroup of integer unimodular matrices $M$ that satisfy (\ref{point_group_eq}).
\end{defin}
It follows that the matrix representation of $\PP$ in  any basis in $\rr^n$ is conjugated to $\GG$ in $GL(n,\rr)$. 

\begin{defin}  Two lattices $\LL$ and $\LL^\prime$ are arithmetically equivalent if the associated lattice groups $\GG$ and $\GG^\prime$ are conjugated in $GL(n,\zz)$, i.e., there exists 
$H\in GL(n,\zz)$ such that:
\begin{equation}
\GG=H^{-1}\GG^\prime H.
\end{equation} 
\end{defin}

Consider now a simple lattice $\LL$ with basis $\{\bse_i\}_{i=1,\dots,n}$ and origin $Q_0$, and $N$ points $Q_1,\dots,Q_N$ not belonging to $\LL$ and not pairwise equivalent modulo $\LL$ (i.e., such that it does not exist a lattice vector $\bst=m^i\bse_i$ ($m^i\in\zz$) such that $Q_\alpha=Q_\beta+\bst$ for $\alpha\ne \beta$, and $\alpha,\beta=0,\dots,N$).

\begin{defin}  A $(N+1)$-lattice with basis $\{\bse_i\}_{i=1,\dots,n}$ is the union of $N+1$ simple lattices $\LL(Q_\alpha,\bse_i)$ with origin at the points
 $Q_0,\dots,Q_N$, i.e., the set 
 \begin{equation}\bigcup_{\alpha=0}^N\{Q_\alpha+m^i\bse_i\, ,\,m^i\in\zz\}.
\end{equation} 
\end{defin} 
The position of the additional points with respect to the origin of the lattice $\LL$, called the skeletal lattice,  is usually described in terms of the shift vectors
\begin{equation}
\label{shift}
\bsp_\alpha=Q_\alpha-Q_0, \qquad\alpha=0,\dots,N.
\end{equation}
Notice that $\bsp_0={\boldsymbol 0}$.

\subsection{The symmetry group of a multilattice}

The symmetric group $S_{N+1}$, viewed  as the group of permutations on the $(N+1)$ points $Q_0,\dots,Q_N$,
\begin{equation}
\label{azione_sigma} 
\sigma: (Q_0,\dots,Q_N)\mapsto(Q_{\sigma(0)},\dots,Q_{\sigma(N)}), \qquad   \sigma\in S_{N+1}
\end{equation}
also acts on the shift vectors in a natural manner.

\begin{pro}
$S_{N+1}$  acts linearly on the $\zz$-module generated by the shift vectors $\{\bsp_1,\dots,\bsp_N\}$ as follows: 
\begin{equation}
\label{action}
\widetilde\sigma :\bsp_\alpha\mapsto 
\bsp_{\sigma(\alpha)} -\bsp_{\sigma(0)}, \qquad\alpha=1,\dots,N,
\end{equation}
for every $\sigma \in S_{N+1}$.
\end{pro}

\noindent\textbf{Proof}.
By (\ref{shift}) and (\ref{azione_sigma}),
\begin{equation*}
\begin{split}
\bsp_\alpha=Q_\alpha-Q_0\mapsto Q_{\sigma(\alpha)} -Q_{\sigma(0)}
&=
Q_{\sigma(\alpha)}-Q_0 -(Q_{\sigma(0)}-Q_0) \\
&= \bsp_{\sigma(\alpha)} -\bsp_{\sigma(0)},
\end{split}
\end{equation*}
which proves the thesis.\hfill $_\blacksquare$
\par\bigskip

We denote by $\SS_{N+1}$ the group of matrices defined by the linear action (\ref{action})
of $S_{N+1}$ on the shift vectors:
\begin{equation}
\label{rapp_b}
\SS_{N+1} =\left\{ (A_\alpha^\beta)\in GL(N,\zz)/ \exists\sigma\in S_{N+1} :A_\alpha^\beta\bsp_\beta=\widetilde\sigma(\bsp_\alpha)=\bsp_{\sigma(\alpha)} -\bsp_{\sigma(0)}
 \right\}.
\end{equation}
 The symmetry of a multilattice can be  described by the set of triples $(\bsR,A_\alpha^\beta,\bst_\alpha)$, 
with $\bsR\in\PP$, $(A_\alpha^\beta)\in\SS_{N+1}$ and $\bst_\alpha\in\LL$ for $\alpha=1,\dots,N$, such that the action of $\bsR$ on the shift vectors is a permutation of the additional points, modulo translations of the lattice: i.e.,
\begin{equation}
\label{master_1}
\bsR\bsp_\alpha=A_\alpha^\beta\bsp_\beta+\bst_\alpha,
\qquad 	\alpha=1,\dots,N.
\end{equation}
Granted (\ref{point_group_eq}), and writing\footnote{$\MM(n\times N,\rr)$ and $\MM(n\times N,\zz)$ are the vector space and $\zz$-module of $n\times N$ real and integer matrices, respectively.} $\bsp_\alpha=P_\alpha^i\bse_i$ and $
\bst_\alpha=T_\alpha^i\bse_i$, with $P_\alpha^i\in\MM(n\times N,\rr)$ $T_\alpha^i\in \MM(n\times N,\zz)$,
 we may rewrite  (\ref{master_1}) in the form
\begin{equation}
\label{master_2}
M_j^iP^j_\alpha=P^i_\beta A_\alpha^\beta+T^i_\alpha,
\qquad 	\alpha=1,\dots,N,
\end{equation}
i.e., letting $M=(M_i^j)$,  $P=(P_\alpha^i)$,  $A=(A_\alpha^\beta)$, $T=(T_\alpha^i)$, 
\begin{equation}
\label{master_3}
MP=PA+T.
\end{equation}
We refer to (\ref{master_2}) or (\ref{master_3}) as the 'master equation'.  We now show that  (\ref{master_3})  defines indeed  a group, which completely identifies the symmetry of the multilattice. 

\begin{pro}\label{proposition-group} Let $P=(P_\alpha^i)$ be given, and let $\
\HH$ be the subset of $\GG\triangleleft GL(n,\zz)$ of matrices $M=(M_j^i)$ such that there exist $A=(A_\alpha^i)\in \SS_{N+1}$ and $T=(T_\alpha^i)\in \MM(n\times N,\zz)$ that satisfy the master equation  (\ref{master_3}). Then 
\par\noindent
i) $\HH$ is a group;
\par\noindent
ii) the map $\HH\to \SS_{N+1}$ defined by (\ref{master_3}) is a group morphism that defines a permutation representation of $\HH$ on the set $\{Q_0,\dots,Q_N\}$, through the relation $A_\alpha^\beta\bsp_\beta=Q_{\sigma(\alpha)}-
Q_{\sigma(0)}$.
\end{pro}

\textbf{Proof}. Both statements follow from the following argument. If $M,H\in \HH$, there exist $A_M,A_H\in \SS_{N+1}$ and $T_M,T_H$ integer matrices, such that 
$$
MP=PA_M+T_M,\qquad HP=PA_H+T_H.
$$
Hence, 
$$
(MH)P=M(PA_H+T_H)=(PA_M+T_M)A_H +MT_H=
P(A_MA_H)+(T_MA_H +MT_H).
$$
Further, by multiplying $MP=PA_M+T_M$ to the left by $M^{-1}$ and to the right by $A_M^{-1}$, we find
$$
M^{-1}P=PA_M^{-1}-M^{-1}T_MA_M^{-1},
$$
and the thesis is proved, since $T_MA_H +MT_H$ and $M^{-1}T_MA_M^{-1}$ are matrices of integers.\hfill $_\blacksquare$
\par\bigskip
We denote by $ \Gamma_{n,N}$ the set of matrices  in $GL(n+N,\zz)$ defined by
\begin{equation}
\label{gamma}
\begin{split}\Gamma_{n,N}=&
\left\{ 
 \left(
\begin{array}{cc}
H&E\\
0&B
\end{array}
\right)\in GL(n+N,\zz)\;/\;
H\in GL(n,\zz),
\right. \\&
\left.
\qquad
\qquad
\qquad
\qquad
\qquad
E\in
\MM(n\times N,\zz),
B\in\SS_{N+1}
 \right\}.
\end{split}
\end{equation}

By the above proposition the symmetry group of a $(N+1)$-lattice can then be characterized by the group of matrices  $\KK\subset \Gamma_{n,N}$ of the form 
\begin{equation}
\label{extended_matrices}
\KK=\left\{
\left(
\begin{array}{cc}
M&T\\
0&A
\end{array}
\right)
\in \Gamma_{n,N}\, / \,M\in\HH,\,\,
MP=PA+T
\right\},
\end{equation}
which we shall refer to as the symmetry group of the multilattice. 

\begin{defin}\label{equivalent_multilattices}
Two $(N+1)$-lattices with shift vectors $P$ and $P^\prime$, and corresponding symmetry groups $\KK$ and $\KK^\prime$  are arithmetically equivalent if $\KK$ and $\KK^\prime$ are conjugated in the set of matrices $\Gamma_{n,N}$, i.e., if there exists a matrix $Q\in\Gamma_{n,N}$ such that
\begin{equation}\label{conjugacy1}
\KK^\prime=Q^{-1}\KK Q.
\end{equation}
\end{defin}

Clearly, $\HH$ is isomorphic to $\KK$ but, in general, there are many inequivalent  multilattices, with different symmetry groups, all isomorphic to the same point group $\HH$ (cf. for instance, \cite{pitteri-zanzotto}).

\section{The master equation as a system of linear equations}

The master equation can be viewed either as a relation that uniquely characterizes the symmetry group $\KK$ of a multilattice, given the shift vectors $\bsp_\alpha$, or as an equation in the unknowns $\bsp_\alpha$, that allows to determine all multilattices that have a given symmetry group $\KK$.   In this paper we take the second point of view, and assume that $\KK$, or rather $\HH$, is given. Specifically, the problem we want to solve is:

\begin{itemize}

\item fix a simple lattice $\LL(Q_0,\bse_i)$  with point group $\PP$ and lattice group $\GG$;

\item choose the number $N$ of additional points $(Q_1,\dots,Q_N)$;

\item fix  a subgroup $\HH\subset\GG$ of the lattice group of the crystal;

\item choose a permutation representation $\HH\to S_{N+1}$ that associates to each $M\in\HH$ a permutation of the points $(Q_0,\dots,Q_N)$;

\item determine the resulting linear representation $\HH\to \SS_{N+1}$, in terms of the shift vectors;

\item solve the master equation (\ref{master_3})  in the unknowns $P$ 
for every $M\in\HH$, and corresponding $A\in\SS_{N+1}$, and for every possible $T$.

\item compare the solutions for different choices of $T$.

\end{itemize}

\par\noindent{\bf Remarks}
\par\medskip\noindent
(i) It is  enough to look for solutions of (\ref{master_2}) such that $P_\alpha^i\in[0,1)$, i.e., in the unit cell of the lattice, defined by \begin{equation}
\label{cella}
\mathcal{C}(\bse_i)=\left\{\bsp=P^i\bse_i / P^i \in[0,1)   \right\}
\end{equation}
for a given lattice basis $\bse_i$. In fact,  given a solution  $(P_\alpha^i)$ of (\ref{master_2}), and  $S_\alpha^i\in\MM(n\times N,\zz)$, then $T_\alpha^{\prime i}:=T^i_\alpha+M_j^iS^j_\alpha-S^i_\beta A_\alpha^\beta\in\MM(n\times N,\zz)$ so that
 $(P_\alpha^i+S_\alpha^i)$ is still a solution of (\ref{master_2}) with $T_\alpha^i$ replaced by $T_\alpha^{\prime i}$. The vectors ${\bsp}'_\alpha=\bsp_\alpha+S_\alpha^i\bse_i$ are translations of the shift vectors  $\bsp_\alpha$ by lattice vectors, and thus are equivalent  modulo the lattice.
 
 \par\medskip\noindent
(ii) Since $\HH$ and $\KK$ are finite, they admit a finite set of generators $(M^{(1)},\dots,M^{(K)})$ and 
$(G^{(1)},\dots,G^{(K)})$, with $G^{(k)}=\left(\begin{array}{cc}M^{(k)}& T^{(k)}\\
0&A^{(k)}\end{array}\right)$; Proposition \ref{proposition-group} allows to conclude that if the master equation holds for each generator, then it holds for all elements of the group $\KK$. Hence (\ref{master_2}), that should hold for every element $G$ of $\KK$, can be replaced by 
\begin{equation}\label{master_generators}
M^{(k)}P-PA^{(k)}=T^{(k)},\qquad k=1,\dots,K.
\end{equation}

\subsection{The master equation as a linear system}

Consider first  the master equation  (\ref{master_2}) for a fixed element $G=\left(\begin{array}{cc}M& T \\ 0&A\end{array}\right)\in\KK$:
it can be rewritten as a conventional system of linear equations. To do so, given $\alpha\in\{1,\dots,N\}$ and $i\in\{1,\dots,n\}$, define
\begin{equation}
a=i+(\alpha-1)n,
\label{indices_0}
\end{equation}
so that $a$ takes values in $\{1,\dots,nN\}$. 
  Conversely, let $a=1,\dots,nN$ and define $\alpha$ and $i$ through the identities
\begin{equation}\label{indices_1}
\alpha=\left[\frac{a-1}{n}\right]+1,\qquad
i=a-(\alpha-1)n,
\end{equation}
where $\left[\cdot\right]$ denotes the integer part of its argument. It is clear that as $a$ varies in $\{1,\dots,nN\}$, then $\alpha$ and $i$ take values in $\{1,\dots,N\}$ and $\{1,\dots,n\}$ respectively.  Let 
\begin{equation}
\label{matricione_L}
\widetilde L^a_{b}:=\delta_\alpha^{\beta} M^i_{j}-\delta^i_{j}A^\beta_{\alpha},
\end{equation}
and 
\begin{equation}\label{XP}
\widetilde P^b:=P^j_\beta,  \qquad \widetilde T^a:=T^i_\alpha,
\end{equation}
where $\alpha$, $i$ are defined as in (\ref{indices_1}) and, for $b\in\{1,\dots,nN\}$
\begin{equation}\label{bbetaj}
\beta=\left[\frac{b-1}{n}\right]+1,\qquad 
j=b-(\beta-1)n,
\end{equation}
with $\delta_\alpha^{\beta}$ and $\delta^i_{j}$ Kronecker deltas. The $nN$-dimensional vector $(\widetilde  P^b)$ has components that are obtained by ordering of the vectors $\bsp_\alpha$. 

In terms of the vectors $\widetilde P$ and $\widetilde T$ and the matrix $\widetilde L$,  the master equation (\ref{master_2}) takes the simple form 
\begin{equation}
\label{linear1}
\widetilde L^a_b\widetilde P^b=\widetilde T^a.
\end{equation}
The above assertion follows from a simple argument: let $\widetilde Y^b:=Y^j_\beta$, with $b=1,\dots,nN$,  $j=1,\dots,n$ and $\beta=1,\dots,N$ consistent with the indexing rule above. Then
\begin{eqnarray*}
\sum_{b=1}^{nN}\widetilde Y^b&=& \widetilde Y^1+\dots+\widetilde Y^n+
 \widetilde Y^{1+(2-1)n}+\dots+\widetilde Y^{n+(2-1)n}+
\dots
\\&&+
\widetilde Y^{1+(N-1)n}+\dots+\widetilde Y^{n+(N-1)n}
\\ 
&=& Y^1_1+\dots+Y_1^n+
 Y_2^{1}+\dots+Y_2^{n}+
\dots
+
Y_N^{1}+\dots+Y_N^{n}
\\ 
&=& \sum_{\beta=1}^N\sum_{j=1}^n Y_\beta^j.
\end{eqnarray*}
Hence
\begin{eqnarray*}
\sum_{j=1}^n M^i_jP^j_\alpha
- \sum_{\beta=1}^{N} P^i_\beta A^\beta_\alpha
=\sum_{\beta=1}^{N} \sum_{j=1}^n 
(M^i_j\delta_\alpha^\beta+\delta_j^iA_\alpha^\beta)P^j_\beta
=
\sum_{b=1}^{nN}\widetilde L^a_b\widetilde P^b.\end{eqnarray*}
Consider now the system of master equations (\ref{master_generators}) for the full set of generators of $\KK$, i.e.,
\begin{equation}\label{master_generators2}
M_j^{(k)i}P^j_\alpha-P_\beta^iA^{(k)\beta}_\alpha=T^{(k)i}_\alpha\qquad\qquad k=1,\dots,K,
\end{equation} 
with $K$ the number of generators of $\KK$.  The associated system of linear equations (\ref{linear1}) is now replaced by a system of the form
\begin{equation}
\label{linear2}
\widetilde L^J_b\widetilde P^b=\widetilde T^J,
\end{equation}
with 
\begin{equation}
\label{linear3}
\widetilde L^J_{b}:=\delta_\alpha^{\beta} M^{(k)i}_{j}-\delta^i_{j}A^{(k)\beta}_{\alpha},\quad
\widetilde P^b=P^j_\beta,
\quad \widetilde T^J=T^{(k)i}_\alpha,
\end{equation}
with $J=1,\dots,nNK$ given by 
\begin{equation}
J=i+(k-1)nN+(\alpha-1)n,
\end{equation}
with inverse
\begin{equation*}
\left\{
\begin{array}{l}
\displaystyle k=\left[ \frac{J-1}{nN}   \right]+1,\\
\displaystyle \alpha=\left[\frac{J-(k-1)nN-1}{n}\right]+1,\\
\displaystyle i=J-(k-1)nN-(\alpha-1)n,
\end{array}
\right.
\end{equation*}
and the relations between $b$, $\beta$ and $j$ are as in (\ref{bbetaj}).

\subsection{Tensor form of the master equation}

The relation between the master equation and the matrix $L$  can be rewritten in more compact form as follows. For $M\in GL(n,\zz)$ and $A\in GL(N,\zz)$, consider the fourth order tensor
\begin{equation}
\label{tensors}
M\otimes A^\top, 
\end{equation}
with components  $(M\times A^\top)^{i\alpha}_{j\beta}=M_j^iA_\alpha^\beta$. It is easy to verify that the set of tensors of the form (\ref{tensors}) is a group with product
\begin{equation}
\label{prodotto}
(M\otimes A^\top)*(H\otimes B^\top)=MH\otimes A^\top B^\top.
\end{equation}
Further, the indexing rules (\ref{indices_0}) and (\ref{indices_1}) define a morphism  between the group of such tensors and the group of invertible $nN\times nN$ matrices.

\begin{pro}
For $i,j=1,\dots,n$, $\alpha,\beta=1,\dots,N$, let 
\begin{equation}
\label{indici1}
a = i+(\alpha-1)n,\qquad b = j+(\beta-1)n.
\end{equation}
then the association rule
\begin{equation}
\label{association}
\widetilde L^a_b:=M^i_jA^\beta_\alpha
\end{equation}
defines a map $M\otimes A^\top\mapsto \widetilde L$ between $GL(n,\rr)\otimes GL(N,\rr)$, with product $*$,  and $GL(nN,\rr)$ which  is a group morphism.
\end{pro}

\noindent\textbf{Proof}. Notice first that  if $M$ and $A$ are invertible, then $\widetilde L$ is invertible, with inverse $\widetilde L^{-1}$ associated to the tensor  $ M^{-1}\otimes A^{-\top}$.
Let now $\widetilde R^a_b:=H^i_jB^\beta_\alpha$:  then 
\begin{equation}
\label{somme}
\begin{split}
\sum_{c=1}^{nN} \widetilde L^a_c \widetilde R^c_b 
&= \sum_{h=1}^{n}\sum_{\gamma=1}^{N} \widetilde L^{i+(\alpha-1)n}_{ h+(\gamma-1)n}\widetilde R^{h+(\gamma-1)n}_{ j+(\beta-1)n}
= \sum_{h=1}^{n}\sum_{\gamma=1}^{N}  M^i_h A^\gamma_\alpha   H^h_j B^\beta_\gamma \\
&
=(M H)^i_j (BA)^\beta_\alpha
=(M H\otimes (BA)^\top)^{i\alpha}_{j\beta}
=[(M \otimes A^\top)*( H\otimes B^\top) ]^{i\alpha}_{j\beta},
\end{split}
\end{equation}
which proves the assertion. \hfill $_\blacksquare$

\par\medskip
Further, the tensors of the form 
(\ref{tensors})  act linearly on the space of real matrices $\MM(n\times N,\rr)$ as follows:
\begin{equation}
\label{somme1}
(M\otimes A^\top):P\mapsto MPA,\qquad P\in\MM(n\times N,\rr).
\end{equation}
Letting $\widetilde P\in\rr^{nN}$ be given by (\ref{XP}), the above action corresponds to the linear action of $GL(nN,\rr)$ on $\rr^{nN}$.
In fact,
\[
\begin{split}
\sum_{b=1}^{nN} \widetilde L^a_b  \widetilde P^b 
=&\sum_{j=1}^{n}\sum_{\beta=1}^{N} \widetilde L^{i+(\alpha-1)n}_{ j+(\beta-1)n}\widetilde P^{ j+(\beta-1)n}=\sum_{j=1}^{n}\sum_{\beta=1}^{N} M^i_jA^\beta_\alpha P^j_\beta = (MPA)^i_\alpha.
\end{split}
\]

\section{The solution procedure}

Before discussing (\ref{linear2}), we start with a simple remark on diagonal systems. 

Consider first a diagonal system of linear equations with integer coefficients
\begin{equation}
\label{diagonal}
DX=S,
\end{equation}
with $D\in{\mathcal M}(l\times m,\zz)$ ($l\geq m$) and $D^J_i=0$ for $J\ne i$, $X\in\rr^m$ and $S\in\zz^l$, i.e.,
\begin{equation}
\left\{
\begin{array}{lll}
D^i_iX^i=S^i & \text{for}& i \leq r \quad (\text{no sum on}\; i)\\
0=S^i &\text{for} & i > r
\end{array}
\right.
\end{equation}
with $r=\text{rank\,}D$.
Define 
\begin{equation}
\label{Sset}
\mathcal{S}=\{(S^1,\dots,S^l)\in\zz^l / 0\leq S^i <D^i_i \text{\;for\;} i =1,\dots,r, S^i=0 \text{\;for\;}i=r+1,\dots,l \}
\end{equation}
and
 \begin{equation}
\label{Bset}
\mathcal{B}^i=\left\{0,\frac1{D_i^i},\frac2{D_i^i},\dots,\frac{D_i^i-1}{D_i^i} \right\} \text{\;for\,} i =1,\dots,r.
\end{equation}
Then the solutions $X$ of (\ref{diagonal}), for $S\in\mathcal{S}$, belong to the set
 \begin{equation}
 \label{solution}
 \begin{split}
\mathcal{X}=&\left\{  (X^1,\dots,X^m) / X^i\in\mathcal{B}^i  \text{\;for\,} i =1,\dots,r ,\right.
\\
&\left.\quad
\text{and}\,X^j\in[0,1)\, \text{for\,} j=r+1,\dots,m \right\}.
\end{split}
\end{equation}
Notice that $S^{r+1},\dots,S^l$ must be zero for the resolubility condition of (\ref{diagonal}) to hold,
and we  have assumed that $X^j\in[0,1)$ for $j=r+1,\dots,m$ without loss of generality.

The following proposition is a consequence of this simple fact:  

\begin{pro}\label{diagonal1} The solutions $X$ of (\ref{diagonal}), with $S\in\zz^r\times\{0\}^{l-r}$, belong to $\mathcal{X}\mod\zz^m$, i.e., 
there exists $Y\in\mathcal{X}$ and $K\in\zz^m$ such that $X=K+Y$.
\end{pro}
\textbf{Proof.} Given $S\in\zz^r\times\{0\}^{l-r}$, then for all $i=1,\dots,r$ there exist $K_i\in\zz$ and $C_i\in\left\{0,1,\dots,D^i_i-1\right\}$ such that (no sum on repeated indices)
$$S^i=D^i_iK_i+C^i.$$
Then $D^i_iX^i=D^i_iK_i+C^i$ (again no sum on repeated indices), and  
$$
X^i=K^i+Y^i \quad\text{with}\quad Y^i\in\mathcal{B}^i,
$$
for $i=1,\dots,r$, and the statement is proved.\hfill $_\blacksquare$
\par\bigskip

Consider now the full system of linear equations (\ref{linear2}): instead of solving it for a fixed value of the right-hand side  $\bst_\alpha$, and hence of $\widetilde T$, we look for solutions for every integer vector $\widetilde T$. 
 Hence, we rewrite (\ref{linear2}) in the form 
\begin{equation}
\label{linear4}
\widetilde L^J_b\widetilde P^b=0\,\mod\zz^{(nNK)},
\end{equation}
with $J=1,\dots,nNK$ and  $b=1,\dots,nN$ and $\widetilde L^J_b$ is  a matrix with integer entries.

It is a classical result that every such matrix can be reduced to a diagonal canonical form, the Smith canonical form \cite{newman}, \cite{gohberg}. Precisely, for every matrix $(\widetilde L^J_b)\in {\mathcal M}(nNK \times nN,\zz)$, there exist matrices
$(U^J_I)\in GL(nNK,\zz)$ and $(V^a_b)\in GL(nN,\zz)$, such that 
\begin{equation}
\label{smith}
\widetilde L^J_b=U^J_I D^I_aV^a_b,\qquad (D^I_a)\in {\mathcal M}(nNK\times nN,\zz),
\end{equation}
with $D^I_a=0$ for $I\ne a$, and $D_i^i$ divides $D_{i+1}^{i+1}$ if $D_{i+1}^{i+1}\ne0$. The Smith canonical form $D$ is unique, whereas the matrices $U$ and $V$ are not. 

Define now:
$$
\mathcal{P}=\{\widetilde P\in\rr^{nN} / \widetilde P= V^{-1}X \text{ with }X\in\mathcal{X}   \}
$$
where $\mathcal{X}$ is defined as in (\ref{solution}) with $m=nN$.

\begin{pro}\label{solutions} Let $\widetilde L\in \MM(nNK\times nN,\zz)$, and $D$ its Smith normal form, with $r=\text{rank}(D)$: then the solutions of (\ref{linear4}) belong to $\mathcal{P}\mod\zz^{nN}$. More precisely, 
there exist exactly $\infty^{Nn-r}\times (D_1^1D_2^2\dots D^r_r)$ solutions  modulo $\zz^{nN}$ of the system (\ref{linear4}), and are given by 
\begin{equation}
\label{linear8}
\widetilde P^b=\overline{V}^b_aX^a,\qquad (X^a)=\left(
\frac {k_1}{D_1^1},\frac {k_2}{D_2^2},\dots,
\frac {k_r}{D_r^r},t_1,\dots,t_{nN-r}\right),
\end{equation}
with $\overline V =V^{-1}$ and 
$$
k_i\in\{0,1,\dots,D_i^i-1\},\quad t_j\in [0,1).
$$
\end{pro}

\par\noindent
\textbf{Proof}. The general procedure to solve (\ref{linear4}) is as follows:  let
$$
X^a=V^a_b\widetilde P^b,
$$
so that, since $(U^J_I)\in GL(nNK,\zz)$,   the system (\ref{linear4}) can be written in the form 
\begin{equation}
\label{linear5}
D^J_aX^a=0\,\mod\zz,
\end{equation}
i.e., 
\begin{equation}
\label{linear6}
\left\{
\begin{array}{l}
D_1^1X^1=0\,\mod\zz,\\
D_2^2X^2=0\,\mod\zz,\\
\dots\\
 D_r^rX^r=0\,\mod\zz,
\end{array}
\right.\end{equation}
where $r=\text{rank\,}(D^J_a)$. By Proposition \ref{diagonal1}, it is sufficient to solve (\ref{linear6}) in the set $\mathcal{X}$, i.e. in the set $[0,1)^m$: we obtain
\begin{equation}
\label{linear7}
\left\{
\begin{array}{l}
X^1=0,\frac1{D_1^1},\frac2{D_1^1},\dots,\frac{D_1^1-1}{D_1^1}\\
X^2=0,\frac1{D_2^2},\frac2{D_2^2},\dots,\frac{D_2^2-1}{D_2^2}\\
\dots\\
X^r=0,\frac1{D_r^r},\frac2{D_r^r},\dots,\frac{D_r^r-1}{D_r^r}\\
X^{r+1}=t_1\in[0,1),\\
\dots\\
X^{nN}=t_{nN-r}\in[0,1),
\end{array}
\right.
\end{equation}
with $t_i$ real parameters, and this yields (\ref{linear8}).
\hfill $\blacksquare$ 

\par\medskip

Once the $X^a$ and the corresponding $\widetilde P^b$ are computed, the right-hand sides of the master equation (\ref{linear2}) may be determined as follows: first rewrite the diagonalized equation (\ref{linear5}) in the form 
\begin{equation}
\label{linear9}
D^J_aX^a=S^J,
\end{equation}
where $D^J_aX^a$ are known. The solvability condition for (\ref{linear9}) is clearly $S^{r+1}=\dots=S^{nN}=0$, while 
by construction $S^i=k_i$ for $i=1,\dots,r$ (cf. Proposition \ref{solutions}). Once the integer components $S^J$ are determined, $\widetilde T^J=U^J_IS^I$ follows immediately.
Finally, formulas (\ref{linear3})$_{2,3}$ yield the solution in terms of the $P^i_\alpha$ and  $T^{(k)i}_\alpha$.

%%%%%%%%%%%%%%%%%%%
\par\medskip

The Smith normal form $D$ of a given integral matrix $\widetilde L$ is unique, but the transformation matrices $U$ and $V$ such that $\widetilde L=UDV$ are not. Since our procedure is based on evaluating a fixed solution $X$ of $DX=0\,\mod\zz^{nNK}$, and then reconstructing $\widetilde P$ as $\widetilde P=V^{-1}X$, the question of the independence of the solution of $U$ and $V$ arises. We now prove that, for all possible choices of $U$ and $V$, the solutions belong to the same set, modulo translations of the lattice.
To see this, recall that, if $U^*$ and $V^*$ are other transformation matrices such that  $\widetilde L=U^*DV^*$, then $U^*=UW$ and $V^*=ZV$ with $W$ and $Z$ such that $WDZ=D$ (cf. \cite{jaeger}). Let now $\widetilde P^*=(V^*)^{-1}X$. Then
\[
\begin{split}
\widetilde L\widetilde P^*&=U^*(U^*)^{-1}\widetilde L(V^*)^{-1}X=U^*W^{-1}U^{-1}\widetilde LV^{-1}Z^{-1}X\\&=U^*W^{-1}DZ^{-1}X=U^*DX,
\end{split}
\]
and since $U^*$ is unimodular integer,  if $DX=0\,\mod\zz^{nNK}$ then $U^*DX=0\,\mod\zz^{nNK}$. Hence, $\widetilde P^*$ is a solution of the original equation $\widetilde L\widetilde P^*=0\,\mod\zz^{nNK}$.  Then any solution $\widetilde P^*=(V^*)^{-1}X$ belongs to $\mathcal{P}\mod\zz^{nN}$. Infact, since $\widetilde P^*$ is a solution of $\widetilde L\widetilde P^*=0\mod\zz^{nNK}$
\begin{eqnarray*}
0&=&\widetilde L\widetilde P^*=\widetilde L(V^*)^{-1}X=\widetilde LV^{-1}Z^{-1}X
\\
&=&UDVV^{-1}Z^{-1}X=UDZ^{-1}X\mod\zz^{nNK},
\end{eqnarray*}
and since $U$ is unimodular integer $DZ^{-1}X=0\mod\zz^{nNK}$. Then from Proposition  \ref{diagonal1} it follows that $Z^{-1}X\in\mathcal{X}\mod\zz^{nN}$, i.e. there exist $K\in\zz^{nN}$ and ${X'}\in\mathcal{X}$ such that:
$$
Z^{-1}X=K+{X'}\qquad\text{and} \qquad \widetilde P^*=V^{-1}(Z^{-1}X)=V^{-1}K+V^{-1}{X'},
$$
so that $\widetilde P^*\in\mathcal{P}\mod\zz^{nN}$.

\section{Arithmetic equivalence}

The following proposition shows that two equivalent multilattices have the same Smith normal form, and provides a basis for a criterion to establish whether two multilattices are equivalent.

Consider two equivalent $(N+1)$-lattices. By definition, their symmetry groups $\KK$ and $\KK^\prime$ are conjugated by some $Q\in\Gamma_{n,N}$ of the form
\begin{equation}
Q=\left(\begin{array}{cc}
H&R\\0&B
\end{array}\right).
\label{coniugante}
\end{equation}
To simplify things, we assume that the generators $(G^{(1)},\dots,G^{(K)})\subset \KK$ and $(G'^{(1)},\dots,G'^{(K)})\subset \KK^\prime$ are chosen to be pairwise conjugate, i.e.,  $G'^{(k)}=Q^{-1}G^{(k)}Q$ for each $k=1,\dots,K$. We write the master equations corresponding to each multilattice as usual (cf. (\ref{linear3})),
$$
\widetilde L\widetilde P=\widetilde T,\qquad \widetilde L'\widetilde P'=\widetilde T',
$$
with reduced form
$$
DX=S,\qquad D'X'=S'.
$$
Also, for a given square matrix $\widetilde W\in \MM(nN\times nN,\zz)$, we denote by $\widetilde W_K\in \MM(nNK\times nNK,\zz)$ the square matrix of the form 
\begin{equation*}
\widetilde W_K=\left.\left(
\begin{array}{cccc}
 \widetilde W & 0 &\dots&0  \\
  0 &  \widetilde W&\dots&0 \\
 0 &  \dots&\dots&0 \\
   0 &  0&\dots&\widetilde W
\end{array}
\right)\right\}\text{$K$ times}.
\end{equation*}

\begin{pro}\label{equivalence_smith} Given two equivalent $(N+1)$-lattices as above, 
$\widetilde L$ and  $\widetilde L^\prime$  satisfy the relation
\begin{equation}
\label{burubu}
\widetilde L^\prime= \widetilde W_K^{-1} \widetilde L \widetilde W,
\end{equation}
where $\widetilde W\in GL(nN,\zz)$ is the integer matrix associated to $H\otimes B^{-\top}$ in (\ref{coniugante}) through the relation (\ref{association}). 
Hence,  the matrices $\widetilde L$ and $\widetilde L^\prime$ have the same Smith normal form:
\begin{equation}
\label{burubu2}
D^\prime= D.
\end{equation}
Further, the vectors $S,S'\in\zz^{nNK}$ are related through
\begin{equation}
\label{burubu3}
S^\prime= U'^{-1}\widetilde W_K^{-1}US+D^\prime Z,
\end{equation}
with 
$U,U'\in GL(nNK,\zz)$ are such that $\widetilde L=UDV$ and $\widetilde L'=U'D'V'$, and $Z\in\zz^{nN}$  is a vector of integers. 
\end{pro}

\textbf{Proof}.  Consider two mutually conjugated generators of $\KK$ and $\KK'$,
$$
G^{(k)}=\left(\begin{array}{cc}
M^{(k)}&T^{(k)}\\0&A^{(k)}
\end{array}\right),\qquad
G'^{(k)}=\left(\begin{array}{cc}
M'^{(k)}&T'^{(k)}\\0&A'^{(k)}
\end{array}\right);
$$
by hypothesis, 
\begin{equation}
G'^{(k)}=Q^{-1}G^{(k)}Q,
\label{coniugante2}
\end{equation}
with $Q$ given by (\ref{coniugante}),
so that, in particular, $M'^{(k)}=H^{-1}M^{(k)}H$ and $A'^{(k)}=B^{-1}A^{(k)}B$. Letting 
$$
L^{(k)}=M^{(k)}\otimes I_N-I_n\otimes (A^{(k)})^\top,\quad
L'^{(k)}=M'^{(k)}\otimes I_N-I_n\otimes (A'^{(k)})^\top,\quad
W= H\otimes B^{-\top},
$$
then
$$
 L'^{(k)}= W^{-1}* L^{(k)}* W.
$$
In fact, by (\ref{prodotto}),
\begin{equation*}
\begin{split}
&(H^{-1}\otimes B^{\top})*(M^{(k)}\otimes I_N-I_n\otimes (A^{(k)})^\top)*(H\otimes B^{-\top})
\\
&\qquad\qquad\qquad=H^{-1}M^{(k)}H\otimes B^{\top}I_NB^{-\top} -H^{-1}I_nH\otimes B^{\top}(A^{(k)})^{\top} B^{-\top}
\\& \qquad\qquad\qquad=M'^{(k)}\otimes I_N-I_n\otimes (A'^{(k)})^\top.
\end{split}
\end{equation*} 
Assertion (\ref{burubu}) then follows by letting $\widetilde W$ be the matrix in $GL(nN,\zz)$ associated to $W$ through the rule (\ref{association}),  and using the definition (\ref{linear3}) of $\widetilde L$.  

Further, for each $k$, (\ref{somme1}) and (\ref{coniugante2}) imply that  
$$
T'^{(k)}=H^{-1}T^{(k)}B +(M'^{(k)}\otimes I_N-I_n\otimes (A'^{(k)})^\top)H^{-1}R,
$$
which in turn means that 
$$
\widetilde T'=\widetilde W_K^{-1}\widetilde T +\widetilde L'J,
$$
with $J\in GL(nN,\zz)$ the integer vector associated to the matrix $H^{-1}R$:  $J$ is integer by consequence of the fact that $H\in GL(n,\zz)$  and $R\in\MM(n\times N,\zz)$.  Finally, we obtain (\ref{burubu3}) by multiplying the above identity by $U'^{-1}$.
 \hfill $_\blacksquare$

\par\bigskip
When $\KK$ has one generator, $\widetilde W_K=\widetilde W$ in the above proposition, which reduces to the statement that for equivalent multilattices $\widetilde L$ and $\widetilde L^\prime$ are conjugated. The converse of Proposition \ref{equivalence_smith} is not true:  as it will be shown later, there are multilattices that have the same associated Smith normal form and that are not equivalent.

%%%%%%%%%%%%%

Proposition \ref{equivalence_smith} has a partial converse. 

\begin{pro}\label{equivalence_smith2} Given two $(N+1)$-lattices as above, assume that their lattice groups $\HH$ and $\HH^\prime$, as well as their images in $\SS_{N+1}$, are conjugated, i.e.,  there exists $H\in GL(n,\zz)$ and $B\in\SS_{N+1}$ such that, writing
$$
G^{(k)}=\left(\begin{array}{cc}
M^{(k)}&T^{(k)}\\0&A^{(k)}
\end{array}\right),\qquad
G'^{(k)}=\left(\begin{array}{cc}
M'^{(k)}&T'^{(k)}\\0&A'^{(k)}
\end{array}\right),
$$
for the generators of $\KK$ and $\KK^\prime$ respectively, then 
\begin{equation}
\label{burubu0}
M'^{(k)}=H^{-1}M^{(k)}H\qquad\text{ and }\qquad A'^{(k)}=B^{-1}A^{(k)}B,
\end{equation}
for each $k=1,\dots,K$. 
If there exists an integer vector $Z\in \zz^{nN}$ such that 
\begin{equation}
\label{burubu4}
S^\prime= U'^{-1}\widetilde W_K^{-1}US+D^\prime Z,
\end{equation}
with the same notations of Proposition (\ref{equivalence_smith}), the two multilattices are equivalent.
\end{pro}

\textbf{Proof}.  Clearly, (\ref{burubu4}) implies that there exists $R\in\MM(n\times N,\zz)$ such that 
$$
\widetilde T'=\widetilde W_K^{-1}\widetilde T +\widetilde L'J,
$$
with $J\in GL(nN,\zz)$ the integer vector associated to the matrix $H^{-1}R$ which, together with (\ref{burubu0}),  implies in turn that 
$$
G'^{(k)}=Q^{-1}G^{(k)}Q,
$$
holds for each $k$, with $Q$ given by (\ref{coniugante}).
 \hfill $_\blacksquare$

\section{Applications to the arithmetic  classification of multilattices}

The above algorithm can help  to solve a major problem of the arithmetic classification of multilattices, namely to generate all equivalence classes of  $(N+1)$-lattices. 
In fact, assume that a subgroup $\HH$ of the lattice group of a Bravais lattice is given. For given $N$, there is only a finite number of inequivalent permutation representations of $\HH$ on $N+1$ elements. Assume that one of these has been chosen, so that a set of generators $M^{k}$ of $\HH$, together with the corresponding $A^{(k)}$, are known. This allows to construct the matrix $L$, and Proposition \ref{solutions}  yields the shift vectors $P$ and the corresponding translations $T$.  To establish which of these solutions are mutually equivalent, we can now to apply Proposition \ref{equivalence_smith2}, with $M'^{(k)}=M^{(k)}$ and $A'^{(k)}=A^{(k)}$. In this case, the conjugants $H$ and $B$ are just normalizers of $(M^{(1)},\dots,M^{(K)})$ and of $(A^{(1)},\dots,A^{(K)})$ respectively. To this purpose, it is necessary to compute the integer matrices $H\in GL(n,\zz)$ such that 
$$
H^{-1}M^{(k)}H=M^{(k)},\qquad \text{for all}\quad k=1,\dots,K ,
$$
and the permutations $B\in \SS_{N+1}$ such that 
$$
B^{-1}A^{(k)}B=A^{(k)},\qquad \text{for all}\quad k=1,\dots,K. 
$$
We are now in the hypothesis of Proposition \ref{equivalence_smith2}, which yields a necessary and sufficient condition for equivalence.

To illustrate the above procedure, we show that, given a subgroup of a lattice group and a permutation representation, our approach yields all  inequivalent multilattices that have that lattice subgroup and permutation representation.
Specifically, focusing on $2$-lattices, we consider for instance the hexagonal point group with trivial permutation representation and show how to obtain the inequivalent  structures 27 and 28 in \cite{fadda_zanzotto_01} .
In this case, $n=3$, $N=1$ and $K=2$.
Consistent with \cite{fadda_zanzotto_01}, we choose as generators of the hexagonal lattice group the matrices 
\[
M^{(1)}=
\left(
\begin{array}{rrr}
  -1 &1&0   \\
-1&0&0   \\
0&0&-1 
\end{array}
\right),
\qquad
M^{(2)}=
\left(
\begin{array}{rrr}
  -1&1&0   \\
0&1&0   \\
0&0&1 
\end{array}
\right),
\]
and $A^{(1)}=A^{(2)}=1$. The matrix $\widetilde L$ corresponding to the master equation  is 
\[
L=\left(
\begin{array}{rrr}
  -2 &1&0   \\
-1&-1&0   \\
0&0&-2 
  \\
-2&1&0
    \\
0&0&0
    \\
    0&0&0  
\end{array}
\right),
\]
with Smith normal form 
\begin{equation*}
D=\left(
\begin{array}{ccc}
  1 &0&0   \\
0&1&0   \\
0&0&6 
  \\
0&0&0
    \\
0&0&0
    \\
0&0&0
\end{array}
\right),
\quad
U=\left( \begin {array}{rrrrrr} 
-2&3&1&0&0&0\\
-1&0&0&0&0&0
\\0&-2&-1&0&0&0
\\-2&3&1&1&0&0\\
0&0&0&0&1&0\\
0&0&0&0&0&1\end {array} 
\right),\quad
V=\left(\begin {array}{rrr} 1&1&0\\0&3&-2\\0&-1&1\end {array}\right).
\end{equation*}
The diagonal system $DX=0\mod \zz^{6}$ has five nontrivial distinct solutions:
$$
X_i= (0,0,i/6),\qquad i=1,\dots,5
$$
and the corresponding shift vectors are 
$$\begin{array}{lll}
P_1= (2/3,1/3,1/2),&
P_2= (1/3,2/3,0),&
P_3= (0,0,1/2),
\\
P_4= (2/3,1/3,0),&
P_5= (1/3,2/3,1/2).&
\end{array}
$$
The translation vectors in the basis that diagonalizes $\widetilde L$ are
$$
S_i=(0,0,i,0,0,0), ,\qquad i=1,\dots,5.
$$
To establish which of these solutions are mutually equivalent, we can now apply Proposition \ref{equivalence_smith2}. To this purpose, it is 
necessary to compute the integer matrices $H\in GL(3,\zz)$ such that 
$$
H^{-1}M^{(1)}H=M^{(1)},\qquad H^{-1}M^{(2)}H=M^{(2)},
$$
and the integers $B$ such that  $BA^{(1)}B=A^{(1)}$ and $BA^{(2)}B=A^{(2)}$, i.e., $B=\pm1$.
A straightforward calculation shows that $H$ must have the form
$$
H=\pm \left(
\begin{array}{ccc}
1&0&0\\
0&1&0\\
0&0&-1
\end{array}
\right),\quad
\pm \left(
\begin{array}{ccc}
1&0&0\\
0&1&0\\
0&0&1
\end{array}
\right).
$$
Now,  $S_1$ and  $S_5$ satisfy (\ref{burubu4}), i.e., 
$$
S_1= U^{-1}\widetilde W_K^{-1}US_5+D^\prime Z,
$$
with  $B=-1$ and $H=I_3$ the identity in $\rr^3$, which implies that  $\widetilde W_K=-I_6$ is the inversion in $\rr^6$. In fact, for this choice, (\ref{burubu4}) reduces to
$$
\left(
\begin{array}{c}
0\\0\\1\\0\\0\\0
\end{array}
\right)
=
-
\left(
\begin{array}{c}
0\\0\\5\\0\\0\\0
\end{array}
\right)
+
\left(
\begin{array}{ccc}
  1 &0&0   \\
0&1&0   \\
0&0&6 
  \\
0&0&0
    \\
0&0&0
    \\
0&0&0
\end{array}
\right)
\left(
\begin{array}{c}
m_1\\m_2\\ m_3\\m_4\\m_5\\m_6
\end{array}
\right)
$$
with $m_i$ integers, which is satisfied by $Z=(0,0,1,0,0,0)$. 

The same argument shows that 
$$
S_2= U^{-1}\widetilde W_K^{-1}US_4+D^\prime Z,
$$
with the same $H$, $B$ and $Z$ as before.

A straightforward check shows that (\ref{burubu4}) cannot hold for other choices of the normalizers.  We  conclude that the $2$-lattices  with shift vectors 
$$ 
P_1= (2/3,1/3,1/2),	\qquad
P_4= (2/3,1/3,0),
$$
i.e., the structures 27 and 28 in \cite{fadda_zanzotto_01} , are the only inequivalent $2$-lattices  with hexagonal point group and trivial permutation representation. 

In fact, the structure corresponding to the shift $P_3$ is not a $2$-lattice: it is just the hexagonal Bravais lattice with half vertical lattice parameter. To see this, it is enough to  notice that, denoting by $(\bse_1,\bse_2,\bse_3)$ the basis of the hexagonal Bravais lattice in which  
the generators of the lattice group $M^{(1)}$ and  $M^{(2)}$ have the integer representation above, $M^{(1)}$ and  $M^{(2)}$  have trivially the same representation in the basis $(\bse_1,\bse_2,\frac12\bse_3)$.

\section*{Acknowledgements} The author acknowledges   valuable discussions with G. Zanzotto and 
P. Cermelli. This work was partial supported by the MATHMAT Project of the University of Padova (Italy).

\end{document}